\documentstyle[times,pramana,epsfig,floats,amstex]{ias}
\makeatletter

\begin{document}
\mark{{}{}}
\title{KPZ Equation in One Dimension and Line Ensembles}

\author{Herbert Spohn}
\address{Physik Department and Zentrum Mathematik, Technische Universit\"{a}t M\"{u}nchen,\\ D-85747 Garching, Germany\\
{\tt spohn@ma.tum.de}}

\abstract{For suitably discretized versions of the Kardar-Parisi-Zhang equation in one space dimension exact scaling functions are available, amongst them the stationary two-point function. We explain one central piece from the technology through which such results are obtained, namely the method of line ensembles with purely entropic repulsion.}

\maketitle
\section{Introduction}

Since STAPHYS 21 our understanding of the statistical properties
of the one-dimensional KPZ equation (Kadar, Parisi, and Zhang
\cite{KPZ}) for surface growth has improved substantially. Thus an
obvious scheme would be to sketch the main results and to list the
relevant articles. While the latter will be done anyhow, it seems
to me that a more interesting strategy is to explain one central
method through which these advances became possible, namely the
mapping to line ensembles. Given the appropriate line ensemble, to
obtain properties of physical interest still requires a tough
asymptotic analysis for which I refer to the articles
\cite{Joh00,Joh03a,Joh03b,SI04,Fer04,FS,RS,NaSa,ImSa}. The link to
multi-matrix models is illustrated in our contribution to the
Proceedings of the ICMP 14 \cite{FPS}.

It seems difficult to handle directly the continuum KPZ equation
and one has to rely on suitable discretizations. In one dimension the slope is locally
conserved. Thus a natural discretization is to assume that the slope takes
only two values and that the dynamics is defind through a suitable stochastic
exchange rule. Thereby one arrives at the totally asymmetric simple exclusion process
(TASEP), which is also a standard model for driven lattice gases in
one dimension \cite{BKS}. We briefly recall the definition. At
each site $j \in \mathbb{Z}$ there is at most one particle. Thus
the occupation variables $\eta_j(t)$ at time $t$ take only the
values 0 (site $j$ empty) and 1 (site $j$ occupied). Under the
stochastic dynamics a particle with an empty right neighbor site
jumps independently to that site after an exponentially
distributed waiting time with rate 1. The master equation for the
Markov chain thus reads
\begin{equation}\label{eq:1}
\frac{d}{dt}f_t (\eta) = L f_t (\eta)\,,\quad t \geq 0\,,
\end{equation}
with generator
\begin{equation}\label{eq:2}
L f (\eta) = \sum_{j\in \mathbb{Z}}
\eta_j(1-\eta_{j+1})\big(f(\eta^{j,j+1})-f(\eta)\big)
\end{equation}
as acting on functions $f$ over configuration space. Here $\eta^{j,j+1}$
denotes the configuration $\eta$ with the occupations at sites $j$
and $j+1$ interchanged.

In the interpretation of surface growth, the height function
$h_j(t)$ at time $t$ has the height differences
$h_{j+1}(t)-h_j(t)=1- 2\eta_{j+1}(t)$ with the convention that
$h_0(t)=2N_t$, $N_t$ being the number of jumps through the bond from 0
to 1 in the time span $[0,t]$. Then our dynamical rule simply
means that local minima of $h_j(t)$ are filled independently after
an exponentially distributed waiting time. The stationary measures of the TASEP
are Bernoulli, $\textrm{Prob}_\rho(\eta_j=1)=\rho$. Therefore
$\langle\eta_i\eta_j\rangle_\rho -\rho^2= \delta_{ij}\rho(1-\rho)$
and the stationary current is $j(\rho)=\rho(1-\rho)$. Since
$j''(\rho)\neq 0$, the TASEP is in the KPZ universality class for
every density $\rho$, $\rho\neq 0,1$. More details can be found in
\cite{PS02a}.

For the initial conditions there are three prototypical choices.
In fact, while the dynamical scaling exponent is always $z=3/2$,
the scaling functions do depend on the initial conditions.\smallskip\\
{\it{- flat initial data}}, which is the standard choice for
numerical simulations. For the TASEP it means
$\eta_j(0)=(1+(-1)^j)/2$. Then $\langle
h_0(t)\rangle=\frac{1}{2}t$. From a statistical mechanics point of
view the object of interest are random fluctuations. Their
behavior is only partially understood. Assuming universality,
there is strong evidence  for
\begin{equation}\label{eq:3}
h_0(t)\cong \frac{1}{2}t - \xi_{\mathrm{GOE}}t^{1/3} \quad
\textrm{for}\; t\to \infty
\end{equation}
with the random amplitude $\xi_{\mathrm{GOE}}$ distributed as the
largest eigenvalue of a $N\times N$ GOE random matrix in the limit
$N\to \infty$ \cite{BR,Fer04}. This distribution is known as Tracy-Widom at $\beta=1$ 
 \cite{TW}. Note that instead of centering we subtracted the
asymptotic mean. In particular, it turns out that
$\langle\xi_{\mathrm{GOE}}\rangle<0$. As a stronger statistical property,
under diffusive scaling  with scale
parameter $t^{2/3}$ and  in the limit $t\to \infty$, the process
\begin{equation}\label{eq:4}
x \mapsto t^{-1/3} \Big( h_{\lfloor t^{2/3}x\rfloor}
(t)-\frac{1}{2}t\Big)\,,\quad \lfloor\cdot\rfloor\, \rm{integer \,
part}\,,
\end{equation}
should have  the same fluctuations as
the top line in Dyson's Brownian motion at $\beta=1$ in the limit
$N\to\infty$, see \cite{Fer04} for a complete discussion.\smallskip\\
{\it{- droplet geometry}}, which takes $\eta_j(0)=1$ for $j\leq 0$
and $\eta_j(0)=0$ for $j\geq0$. In terms of the height,
$h_j(0)=|j|$. The asymptotic shape of the droplet is $h_{\lfloor
xt\rfloor}(t)\cong t(x^2+1)/2$, $|x|\leq 1$, which has nonzero
curvature in contrast to the flat initial conditions. The droplet
geometry occurs naturally in the growth of an Eden cluster and
other growth processes in the plane starting from a point seed. In
our context the droplet geometry derives its fame from the work of
K. Johansson \cite{Joh00}, who proved for the height at the origin
\begin{equation}\label{eq:5}
h_0(t)\cong \frac{1}{2}t - \xi_{\rm{GUE}} t^{1/3} \quad \rm{for}\;
\textit{t}\to \infty
\end{equation}
with the random amplitude $\xi_{\mathrm{GUE}}$ distributed as the
largest eigenvalue of a $N\times N$ GUE random matrix in the limit
$N\to \infty$. For the droplet geometry even stronger property
(\ref{eq:4}) is available. The limit is the stationary Airy
process \cite{PS02c,Joh03a}. (For the sake of notational clarity we
omit numerical multiplicative factors. They are provided in the
references given.)\smallskip\\
{\it{- stationary initial conditions}}. This means that at time
$t=0$ the $\eta_j(0)$ are independent with ${\rm{Prob}}_\rho
(\eta_j(0)=1)=\rho$. The stochastic process $\eta_j(t)$ is 
stationary in both variables and can be viewed as a
two-dimensional field theory in infinite volume. The natural statistical
object is then the two-point function
\begin{equation}\label{eq:6}
S(j,t)=\langle \eta_j (t) \eta_0 (0)\rangle_\rho - \rho^2\,.
\end{equation}
It has the scaling form, introducing the compressibility $\chi(\rho) = \rho(1-\rho)$,
\begin{equation}\label{eq:7}
S(j,t)\cong
\chi(4\chi^{1/3}t^{2/3})^{-1}
 \frac{1}{8}g'' \big(\big(4
\chi^{1/3} t^{2/3})^{-1}(j-(1-2\rho)t)\big)
\end{equation}
for $j-(1-2\rho)t={\mathcal{O}}(t^{2/3})$ and $t\to \infty$. A
plot of the scaling function $g''$ can be found in \cite{PS02b}.
For the TASEP a proof based on line ensembles is under
construction \cite{Sp04}.

\section{Nonintersecting Line Ensembles}

Line ensembles are familiar from bosonic and fermionic paths
integrals. In our context the relevant example are free fermions
on the one-dimensional lattice with Hamiltonian
\begin{equation}\label{eq:8}
H= \sum_j\big(a^\ast(j)a(j+1)+a^\ast(j+1)a(j)\big)\,.
\end{equation}
Here the Fermi field $a(j)$ satisfies the anticommutation
relations $\{a(i),a(j)\}=0$, $\{a(i),a(j)^\ast\}= \delta_{ij}$.
Let us denote by $x_j(s)$, $j=1,\ldots,N$,  a collection of independent
symmetric nearest neighbor random walks on $\mathbb{Z}$ with jump
rate 1. We condition them not to intersect, indicated by NC for
non-crossing constraint. Then the $N$-particle transition
amplitude is expressed as
\begin{eqnarray}\label{eq:9}
&&\hspace{8pt}\langle x_1,\ldots,x_N|e^{-sH}|y_1,\ldots,y_N\rangle =\nonumber\\
&&\hspace{10pt}{\rm{Prob}^{NC}}\big(x_1(s)=x_1,
\ldots,x_N(s)=x_N|x_1(0)=y_1,\ldots,x_N(0)=y_N\big)\,,
\end{eqnarray}
i.e.~the random walks start at $y_1<\ldots<y_N$ at time 0 and end
up at $x_1<\ldots<x_N$ at time $s$ constrained not to intersect.
Since the constraint is purely geometric, it is also called
entropic repulsion. The transition probability of the TASEP,
$e^{Lt}$, cannot be written as a line ensemble with purely
entropic interaction. Nevertheless, in a much more hidden way
there is such a line ensemble, but it depends in a delicate way on
the initial conditions as will be explained in the following
section.

There is a second example which is formulated rather differently,
but close to (\ref{eq:8}) in fact. Following Dyson  \cite{Dys62},
we consider the stochastic
process $s \mapsto A(s)$ which takes values in the $N\times N$
Hermitian matrices and is governed by the linear Langevin equation
\begin{equation}\label{eq:10}
\frac{d}{ds}A(s)=-\frac{1}{N}A(s)+\dot{W}(s)\,.
\end{equation}
Here $W(s)$ is an $N\times N$ Hermitian matrix
whose matrix elements are independent Brownian motions. We assume
$A(s)$ to be stationary in time. $A(s)$ has the real eigenvalues
$\lambda_{-N+1}(s)<\ldots<\lambda_0(s)$. They never cross. At a fixed
time $s$, $A(s)$ has the
Gaussian distribution $Z^{-1}\exp[-\frac{1}{N}{\rm{tr}} A^2]dA$,
known as GUE of random matrices \cite{Me91}. The eigenvalues
$\lambda_j(s)$, $j=-N+1,\ldots,0$, form a line ensemble with
purely entropic repulsion. In particular, the 2-dimensional random
field
\begin{equation}\label{eq:11}
\phi_N(x,s)=\sum^0_{j=-N+1} \delta(\lambda_j(s)-x)
\end{equation}
has  determinantal moments constructed out of the extended Hermite
kernel \cite{TrWi}.

As to be detailed, very roughly, the height function in the growth process has the
same statistics as the top eigenvalue in Dyson's Brownian motion
(\ref{eq:10}), where space $x$ corresponds to the Brownian motion
time $s$ and the growth time $t$ corresponds to the matrix size
$N$. This correspondance explains why the universal distribution
functions for the KPZ equation are expressed through quantities
appearing also in the edge scaling of random matrices.

\section{Line Ensemble for the TASEP}

We pick the case of step initial conditions, which is the easiest
one to explain and well illustrates the principle. Thus initially
$\eta_j=1$ for $j\leq 0$ and $\eta_j=0$ for $j> 0$. Let us label
the particle initially at site $j$ by the index $-j+1$. We define
$w(i,j)$ as the $i$-th jump time of particle number $j$. Thus
$w(i,j)$, $i,j=1,2,\ldots$, is an infinite array of independent
exponentially distributed random variables, Prob$(w(i,j)\in
[y,y+dy])=e^{-y}dy$, $y\geq 0$. Given the $w(i,j)$'s one constructs from
them a growth process by filling the square $(i,j)$ after the
random time $w(i,j)$ starting from the moment when the two
adjacent squares $(i-1,j)$ and $(i,j-1)$ are already filled. Let us denote
by $G(m,n)$ the time when the square $(m,n)$ is first filled. Then
the cluster grown at time $t$, $\{(m,n)|G(m,n)\leq t\}$, has as
its border the height function $h_j(t)$ up to a rotation by
$\pi/4$.

$G(m,n)$ can be considered as a sort of dual growth process. It
satisfies the random recursion relation
\begin{equation}\label{eq:11a}
G(m,n)= \max \{G(m-1,n),G(m,n-1)\}+w (m,n)\,,
\end{equation}
$G(0,n)=0=G(m,0)$. $G(m,n)$ has also the interpretation of the
optimal time for a last passage directed percolation, which can be
seen by introducing the directed polymer, $\omega$. It is an
up/right path on $({\mathbb{Z}}_+)^2$ starting at $(1,1)$. The
``length'' of a polymer is given by
\begin{equation}\label{eq:12}
\tau(\omega)=\sum_{(i,j)\in\omega}w(i,j)\,.
\end{equation}
$G(m,n)$ is the length of the optimal path at fixed endpoints,
\begin{equation}\label{eq:13}
G(m,n)=\max_{\omega:(1,1)\to(m,n)}\tau(\omega)\,.
\end{equation}
Since $h_j(t)$ is increasing linearly in $t$, so is $G(n+k,n-k)$
in $n$, and the scaling behavior of $h_j(t)$ for large $t$ is
determined by the scaling behavior of $G(n+k,n-k)$ for large $n$.
For example, $G(n,n)$ is the time it takes for $h_0(t)$ to reach
the level $2n$. Thus the assertion (\ref{eq:5}) follows from a
corresponding one for $G(n,n)$ in the limit of large $n$.

\begin{figure}
\begin{center}
\epsfig{file=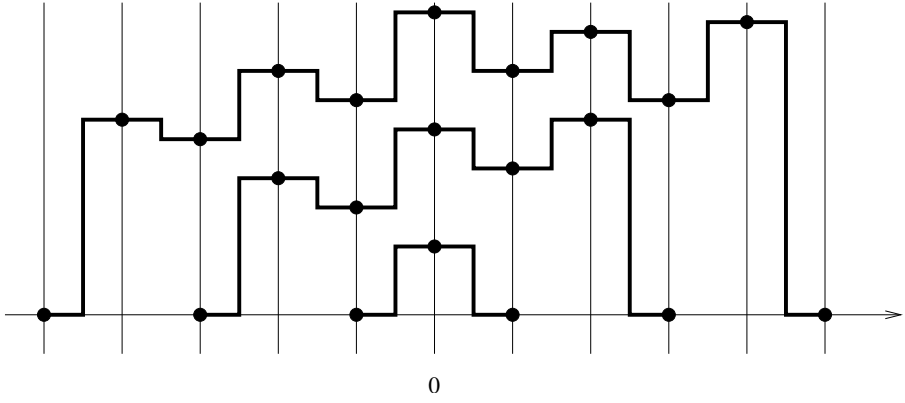,width=12cm}
\end{center}
\caption{Line ensemble for the TASEP with $\tau=3$. The lines
 visualize \newline the
non-crossing constraint.}
\end{figure}

We are now in the position to construct out of $G(m,n)$ an
ensemble of lines with purely entropic repulsion. We fix $\tau$ and
introduce the lines $h_\ell(j)$, $\ell=0,\ldots,-\tau+1$.
$h_\ell(j)\in \mathbb{R}_+$ and $h_\ell(2j)-h_\ell(2j-1)\geq 0$
while $h_\ell(2j+1)-h_\ell(2j)\leq 0$. Furthermore $h_\ell(j)=0$
for $|j|\geq 2\tau-2\ell-1$, see Figure 1. Each jump of absolute size
$\delta$ carries the weight $e^{-\delta/2}$, $\delta\geq 0$. The
weight of the whole line ensemble is then the product of the
weights of all jumps. Finally one has to impose the non-crossing constraint.
It is most easily formulated through extending $j\mapsto h_\ell (j)$
to piecewise constant lines in the plane as
\begin{equation}\label{eq:14}
h_\ell(s)=h_\ell(j)\quad {\rm{for}}\quad j-\frac{1}{2}\leq s\leq
j+\frac{1}{2}\,.
\end{equation}
Note that for each constant piece both endpoints are included. We then require
\begin{equation}\label{eq:15}
{\rm (NC)}\hspace{2cm} h_{\ell-1}(s) < h_\ell(s)\hspace{3cm}
\end{equation}
for all $\ell$, all $s$, and all endpoints, compare with Figure 1.

The idea of the line ensemble is that the top line, $h_0(j)$,
records the statistics of $G(m,n)$, while the others are
introduced only for book-keeping purposes. This can be seen by
generating the line ensemble through the Robinson-Schensted-Knuth
algorithm \cite{Sch61} (it is the Knuth of TeX). $\tau$ is regarded
as growth time parameter and we set $w(1,1)=h_0(0,\tau=1)$, see
Figure 2. The up-step (at $s=-\frac{1}{2}$) is shifted one unit to
the left and the down-step (at $s=\frac{1}{2}$) is shifted one
unit to the right, yielding $\widetilde{h}_0(j,\tau=1.5)$, $j=-1,0,1$.
Then we add mass according to
$h_0(-1,\tau=1.5)=\widetilde{h}_0(-1,\tau=1.5)+w(1,2)$,
$h_0(0,\tau=1.5)=\widetilde{h}_0(0,\tau=1.5)$,
$h_0(1,\tau=1.5)=\widetilde{h}_0(1,\tau=1.5)+w(2,1)$, \textit{etc.}. In this
way the top line $h_0(s,\tau)$, $s\in \mathbb{R}$, $\tau=1,2,\ldots$,
satisfies with the identities
\begin{equation}\label{eq:15a}
G(\tau+j,\tau-j)=h_0(2j,\tau)\,, \; G(\tau+j,\tau-j-1)=h_0(2j+1,\tau)\,.
\end{equation}
The line $h_{-1}$ is generated from $h_0$ through a fully
deterministic rule: When moving the up-steps and down-steps of
$h_0$ to $ \widetilde{h}_0$ there will be excess mass at
coalescing steps. This excess mass is recorded as nucleation for
$h_{-1}$. The up-steps and down-steps of $h_{-1}$ move
deterministically in the same fashion as explained for $h_0$. The
nucleation for the line $h_{-2}$ is generated by the excess mass
of $h_{-1}$, \textit{etc.}. Note that the only random input is to
the top line $h_0$. Of course, it requires an argument, see
\cite{ImSa,Sp04}, that at growth time $\tau$ the lines have exactly the
probability distribution as claimed, namely weighing the jumps
exponentially and imposing the non-crossing constraint.

\begin{figure}
\begin{center}
\epsfig{file=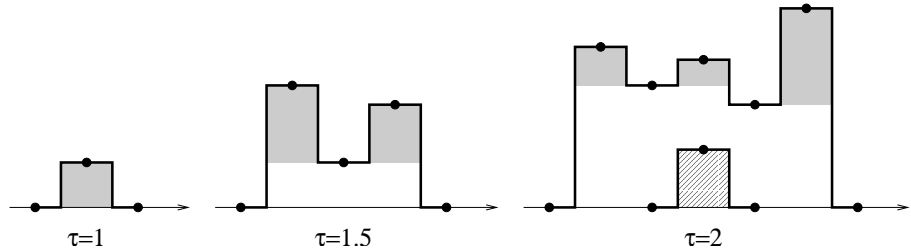,width=12cm}
\end{center}
\caption{The RSK algorithm  for the TASEP. Light gret areas are
mass de-\newline posited according to the growth rule. The shaded
area is the mass annihila-\newline ted in line $0$ and copied to
line $-1$. The lines visualize the updating rule.}
\end{figure}

We have to issue a word of caution. For the multi-line growth
process probabilities referring to two different growth times are
meaningful. In particular we may want to enquire on $h_0(s,\tau)$ for
two different $\tau$'s. This information is no retraceable from the
static, fixed $\tau$, line ensemble. In case of the TASEP, line
ensembles can be used to analyze the statistics of $h_j(t)$, $j\in
\mathbb{Z}$, jointly with prescribed initial conditions but not
for, say, $h_0(t_1)$ and $h_0(t_2)$ jointly.

As in (\ref{eq:11}) for Dyson's Brownian motion the line ensemble
$h_\ell(j,\tau)$ determines the random field $\phi_\tau(x,j)$ through
\begin{equation}\label{eq:16}
\phi_\tau (x,j)=\sum^0_{\ell=-\tau+1} \delta(h_\ell(j,\tau)-x)\,.
\end{equation}
In analogy to femionic path integrals it is convenient to think of $x \in \mathbb{R}_+$ as space,
$j\in \mathbb{Z}$, $|j|\leq \tau$, as time, and $\tau$ as the number of
growth steps. As already anticipated $\phi_\tau$ has determinantal
moments. This means that for a time-ordered sequence $u_1 \leq
\ldots \leq u_m$ and arbitrary space points $x_1,\ldots, x _m$ one
has
\begin{equation}\label{eq:17}
\langle \prod^m_{j=1} \phi_\tau(x_j,u_j)\rangle = \det
\{K^{(2\tau)}(x_i,u_i\,;\;x_j,u_j)\}_{1\leq i,j\leq m}\,.
\end{equation}
To write down a formula for the kernel $K^{(2\tau)}$ we introduce the
operators $T_\pm$ on $L^2(\mathbb{R})$ as given by the integral
kernels, $x,y \in \mathbb{R}$,
\begin{eqnarray}\label{eq:18}
&& T_+ (x,y)= \left\{ \begin{matrix} e^{-(x-y)/2} & , &
{\rm{if}}\; x\geq y\,,
\\ 0 & , & {\rm{if}}\; x<y\,, \end{matrix}\right.\nonumber\\
&& T_- (x,y)= \left\{ \begin{matrix} 0  & , & {\rm{if}}\; x> y\,,
\\ e^{-(y-x)/2} & , & {\rm{if}}\; x\leq y \,.
\end{matrix}\right.
\end{eqnarray}
We also introduce the projection $P_-$ onto the half line
$(-\infty,0]$. Then, for $\tau\geq 1$, $x,y>0$,
\begin{equation}\label{eq:19}
K^{(2\tau)}(x,2u;y,2v)=(T_+T_-)^u(T_+T_-^{-1})^\tau
P_-(T_+^{-1}T_-)^\tau(T_+T_-)^{-v}(x,y)
\end{equation}
in case $-\tau<u\leq v<\tau$ and
\begin{equation}\label{eq:20}
K^{(2\tau)}(x,2u;y,2v)=(T_+T_-)^u((T_+T_-^{-1})^\tau P_-(T_+^{-1}T_-)^\tau
-1)(T_+T_-)^{-v}(x,y)
\end{equation}
in case $-\tau<v<u<\tau$. There are similar formula for odd arguments.

As a particular case one may consider equal times $u,v=0$, i.e.~
the moments of $\phi_{2\tau}(x,0)$. Specializing in (\ref{eq:19})
results in the equality
\begin{equation}\label{eq:21}
K^{(2\tau)}(x,0;y,0)= \sum^{\tau-1}_{j=0} L_j(x)
L_j(y)e^{-x/2}e^{-y/2}\,,
\end{equation}
$x,y>0$, where $L_j$ are the order 0 Laguerre polynomials
normalized as
\begin{equation}\label{21a}
\int^\infty_0 d x e^{-x} L_i(x) L_j(x)= \delta_{ij}\,.
\end{equation}
Such a kernel is known from the theory of random matrices
\cite{Wi}. There one considers $N\times N$ complex matrices $A$ where
the matrix elements $A_{ij}$ are independent Gaussian random
variables normalized as $\langle|A_{ij}|^2\rangle=1$. Then the
eigenvalues $A^\ast A$ have determinantal moments with kernel
(\ref{eq:21}). Thus the eigenvalues of $A^\ast A$ have the same
joint distribution as $h_\ell (0,\tau)$, $\ell=0,\ldots,-\tau+1$.

It follows from (\ref{eq:17}) that the joint distribution function
for the top line at times $u_1,...,u_m$ is expressed through the
determinant of an operator acting on $L^2({\mathbb{R}}) \otimes
{\mathbb{C}}^m$, i.e.~ acting on $m$-spinors over $\mathbb{R}$. We
choose $u_1< \ldots<u_m$ and define the integral kernel
\begin{equation}\label{eq:21b}
{\mathcal{K}}(x,i;y,j)=K^{(2\tau)}(x,u_i;y,u_j)\,.
\end{equation}
Let $\mathcal{G}$ be multiplication by the indicator function
$\chi_{[s_j,\infty)}(v)$, $s_j>0$. Then
\begin{equation}\label{eq:22}
{\rm{Prob}}(h_0(u_1,\tau)\leq s_1,\ldots,h_0(u_m,\tau)\leq s_m)=\det
(1-{\mathcal{G}}^{1/2}{\mathcal{K}} {\mathcal{G}}^{1/2})\,.
\end{equation}
To obtain the asymptotic statistics of $h_0(j,\tau)$ 
thus requires an understanding of the asymptotics
of the kernel $K^{(2\tau)}$ as $\tau\to \infty$ at suitable rescaling of
the arguments $(x,u;y,v)$.

\section{Extensions}

We discussed the line ensemble for step initial conditions of the
TASEP. For large $\tau$, the dual growth process $h_0(j,\tau)$ has a
definite shape, which is bent downwards and has a jump of
macroscopic size at both endpoints. Thus under edge scaling
$h_0(j,\tau)$ is governed by the stationary Airy process, which has
joint distributions of a structure very similar to (\ref{eq:22});
only the kernel $\mathcal{K}$ is replaced by the extended Airy
kernel \cite{Joh03a,PS02c}.

By the same method also general step initial conditions can be
handled, to say: we pick the left density $\rho_-$ and the right
density $\rho_+$ and require that under the initial conditions for
the TASEP the $\eta_j$'s are independent with
Prob$(\eta_j=1)=\rho_-$ for $j\leq 0$ and Prob$(\eta_j=1)=\rho_+$
for $j>0$. In the mapping to the directed polymer the $w(i,j)$'s
have to be extended by the row $w(j,0)$, $j\geq 0$, and the column
$w(0,j)$, $j\geq 0$. The directed polymer starts at $(0,0)$. More
precisely the $w(i,j)$ are independent exponentials with the
following means,
\begin{eqnarray}\label{eq:23}
&&\langle w(i,j)\rangle =1 \hspace{4pt} \textrm{for}  \hspace{4pt}
i,j\geq
1\,, \quad w(0,0) =0\,,\nonumber\\
&&\langle w(j,0)\rangle =(1-\rho_+)^{-1}\,, \quad \langle w(0,j)\rangle =\rho_-^{-1}  \hspace{4pt} \textrm{for}  \hspace{4pt} j>0\,.
\end{eqnarray}

The construction explained in the previous section can be carried
through only if the $w(i,j)$'s are independent exponentials with
Prob$\big(w(i,j)\in
[y,y+dy]\big)=(\alpha_i+\beta_j)\exp[-(\alpha_i+\beta_j)y]dy$,
$y>0$, $\alpha_i,\beta_j>0$. Starting from the left, the weight
for an up-step of size $\delta$ is then $\exp[-\alpha_j\delta]$
and for a down-step of size $\delta$ it is $\exp[-\beta_j\delta]$,
$\delta>0$. For the step initial conditions, $\rho_-=1$,
$\rho_+=0$, clearly
$\overrightarrow{\alpha}=(\frac{1}{2},\frac{1}{2},\ldots)=\overrightarrow{\beta}$.
In the general case, considered here, the obvious choice is
$\overrightarrow{\alpha}=(\frac{1}{2}-\rho_+,\frac{1}{2},\ldots)$
and
$\overrightarrow{\beta}=(\rho_--\frac{1}{2},\frac{1}{2},\ldots)$,
which requires $0\leq \rho_+<\frac{1}{2}$, $\frac{1}{2}<\rho_-\leq
1$. Up to an error of order 1 the TASEP height is then represented by the 
corresponding height of the dual growth process.

The restrictions on $\rho_+$, $\rho_-$ are somewhat surprising,
since the directed polymer with random waiting times according to
(\ref{eq:23}) is well defined for any choice of $\rho_+,
\rho_-$. To cover the left out domain in densities one has to
analytically continue in $\rho_+$, $\rho_-$. The details tend to
be lengthy and we refer to \cite{ImSa} for such a continuation in
case of the discrete time TASEP. Here we  briefly explain 
stationary initial conditions, $\rho_+=\rho=\rho_-$, as the one of
most physical interest. In this case 
$\alpha_0+\beta_0=0$, formally, which means that the weight of the very first
deposition is 1 independent of its height. Such a weight cannot be
normalized to 1, but averages of physical quantities can still be
computed through a suitable limit procedure. As a net result one
finds that for joint distribution functions 
\begin{equation}\label{eq:24}
\textrm{Prob}_\rho\big(h_0(u_1,\tau)\leq s_1,\ldots,h_0(u_m,\tau)\leq
s_m\big)=\frac{d}{ds}\det(1-{\mathcal{G}}_s^{1/2}
{\mathcal{\widetilde{K}}}{\mathcal{G}}_s^{1/2})_{s=0}+
{\mathcal{O}}(1)\,,
\end{equation}
compare with (\ref{eq:22}). Here Prob$_\rho$ refers to the stationary TASEP with initial
Bernoulli measure of density $\rho$. $h_0(u,\tau)$ is the top line of
the corresponding line ensemble after $\tau$ growth steps. ${\mathcal
G}_s$ is the multiplication by the indicator function of the
interval $[s_j+s,\infty)$ and $\mathcal{{\widetilde{K}}}$ is the
extended Laguerre kernel (\ref{eq:19}) perturbed by a rank 1
operator which contains the information on the density $\rho$.
Such a structure seems to be novel.

Step initial conditions lead to boundary terms for the directed
polymer. A further natural modification is to retain the
independent exponentially distributed $w(i,j)$ of mean 1 but to
alter along the diagonal to Prob$(w(i,i)\in [y,y+dy])= \gamma
e^{-\gamma y}d$, $y>0$, $\gamma>0$. This results in two
well-studied, but unresolved problems:\\
\textit{(i)} For the TASEP the rate for jumps from 0 to 1 is
changed from 1 to $\gamma$. For $\gamma\ll 1$, particles pile up
behind the slow bond and the average current is reduced. It is
conjectured that at $\gamma_c=1$ the current reaches its maximal
value of $\frac{1}{4}$ \cite{JaLe}. A more intricate scenario is
proposed in \cite{deN}.\\
\textit{(ii)} On the other hand the directed polymer can be
interpreted as an interface in a random potential with a pinning
potential along the diagonal $\{i=j\}$. For $\gamma<\gamma_c$ the
interface is pinned and it unbinds as $\gamma\to \gamma_c$. For
$\gamma>\gamma_c$ the directed polymer has $n^{2/3}$ transverse
fluctuations \cite{Joh}. It is conjectured that the unbinding transition
occurs at $\gamma_c=1$ \cite{HwNa,TaLy}.

The exponential distribution of the $w(i,j)$'s derives from the
exponential waiting times in the jumps of the TASEP. On the other
hand, from the point of view of the directed polymer any other
distribution looks fine, perhaps imposing some moment conditions.
Unfortunately the construction of purely entropic line ensembles
requires that the $w(i,j)$'s are allowed to be independent with
the geometric distribution Prob$(w(i,j)=n)=(1-a_{ij})(a_{ij})^n$,
$n=0,1,\ldots$, where the $a_{ij}$ must be of product form,
$0<a_{ij}<1$ and $a_{ij}=b_i b_j$. This includes the
\textit{exponential limit}, discussed previously. In the opposite limit, the
\textit{Poisson limit}, the $w(i,j)$'s are mostly 0 and once in
while take the value 1. Thereby one arrives at a Poisson point
process in the quadrant $\{x_1\geq 0, x_2\geq 0\}$ with an average density, 
$\rho$,
of product form, $\rho(x_1,x_2)=\rho_1(x_1)\rho_2(x_2)$. The directed polymer
lives then in the continuum. It moves from Poisson point to
Poisson point, linearly interpolating, restricted to be increasing
in both coordinates and starting from $(0,0)$ reaching some
prescribed endpoint $(x_1,x_2)$ in the positive quadrant. The
corresponding growth process is the polynuclear growth (PNG). It
has a continuous growth time, $t$, and the height function
$h(x,t)$ is over the real line. $x \mapsto h(x,t)$ is piecewise
constant and takes integer values only. The dynamical rules can be
read off the directed polymer. Accordingly, an up-step moves with
speed one to the left and a down-step with speed one to the right.
Whenever two steps meet they annihilate. In addition, randomly and
uniformly in space-time, a spike of height one is nucleated on the
top of the current height profile. The spike, then immediately
widens under the deterministic motion.
The line ensemble for the PNG is constructed as before, namely by
using the mass annihilated at line $h_{-\ell}(x,t)$ as nucleation in
the lower lying line $h_{-\ell-1}(x,t)$.  The
droplet geometry can be imposed by requiring that initially
$h(x,0)=0$ for $x\neq 0$ and $h(0,0)=1$ and that nucleation is
allowed only on the top of the first layer. 
As for the TASEP, the
resulting random field has determinantal moments. 

In general, however, line ensembles do not have to be determinantal. A point
in case is the PNG half droplet with an extra source of rate $\gamma$ at the
origin  \cite{SI04}.  In the corresponding line
ensemble the lines are pinned at $x=t$, i.e. $h_{-\ell}(t,t)=-\ell$,
while at $x=0$ they satisfy a ``pairing rule''. In particular, the
point process $\{h_{-\ell}(0,t),\ell=0,1,\ldots\}$ has its moments given
through a suitable Pfaffian. The top line at $x=0$ has
in the scaling limit for $\gamma=0$ the same fluctuations as the
largest eigenvalue of a GSE random matrix and for $\gamma=1$ the
one of a GOE random matrix.
A further point in case is the PNG model with flat initial
conditions \cite{Fer04}. While in space-time the process is not determinantal,
the point process $\{h_{-\ell}(0,t),\ell=0,1,\ldots\}$ has moments given
through a Pfaffian. \medskip\\
{\bf Acknowledgements}. I benefited a great deal from the collaboration
with Patrik Ferrari and Michael Pr\"{a}hofer.

\end{document}